\documentclass[a4paper,11pt]{article}
\pdfoutput=1 

\usepackage{jcappub} 

\usepackage[T1]{fontenc} 

\title{\boldmath Stellar Interferometry for Gravitational Waves}

\author[a]{I.~H.~Park,}
\author[a]{K.~-Y.~Choi,}
\author[a]{J.~Hwang,}
\author[b]{S.~Jung,}
\author[c]{D.~H.~Kim,}
\author[a]{M.~H.~Kim,}
\author[d]{C.~-H.~Lee,}
\author[a]{K.~H.~Lee,}
\author[e]{S.~H.~Oh,}
\author[f]{M.~-G.~Park,}
\author[g]{S.~C.~Park,}
\author[h,i,j]{A.~Pozanenko,}
\author[k]{C.~D.~Rho,}
\author[a]{N.~Vedenkin,}
\author[l,1]{E. Won\note{Corresponding author.}}

\affiliation[a]{Department of Physics, Sungkyunkwan University (SKKU), Suwon, 16419, Republic of Korea}
\affiliation[b]{Department of Physics and Astronomy, Center for Theoretical Physics, Seoul National University, Seoul, 08826, Republic of Korea}
\affiliation[c]{Department of Physics and Astronomy, Astronomy Program,	Seoul National University, Seoul, 08826, Republic of Korea}
\affiliation[d]{Department of Physics, Pusan National University, Busan 46241, Republic of Korea}
\affiliation[e]{Division of Basic Researches for Industrial Mathematics, National Institute for Mathematical Sciences, Daejeon, 34047, Republic of Korea}
\affiliation[f]{Department of Astronomy and Atmospheric Sciences, Kyungpook National University, Daegu 702-701, Republic of Korea}
\affiliation[g]{Department of Physics and IPAP, Yonsei University, Seoul 03722, Republic of Korea}
\affiliation[h]{Space Research Institute of the Russian Academy of Sciences (IKI), 84/32 Profsoyuznaya Str, Moscow, Russia, 117997}
\affiliation[i]{National Research University Higher School of Economics, Myasnitskaya 20, Moscow, 101000 Russia}
\affiliation[j]{Moscow Institute of Physics and Technology (MIPT), Institutskiy Pereulok, 9, Dolgoprudny, 141701 Russia}
\affiliation[k]{Natural Science Research Institute, University of Seoul, Seoul, 02504, Republic of Korea}
\affiliation[l]{Department of physics, Korea University, Seoul 02841, Republic of Korea}

\emailAdd{ilpark@skku.edu}
\emailAdd{kiyoungchoi@skku.edu}
\emailAdd{jungseek@skku.edu}
\emailAdd{sunghoonj@snu.ac.kr}
\emailAdd{ki1313@yahoo.com}
\emailAdd{vader0210@gmail.com}
\emailAdd{clee@pusan.ac.kr}
\emailAdd{lkh6670@naver.com}
\emailAdd{shoh@nims.re.kr}
\emailAdd{mgp@knu.ac.kr}
\emailAdd{sc.park@yonsei.ac.kr}
\emailAdd{apozanen@iki.rssi.ru}
\emailAdd{cdr397@uos.ac.kr}
\emailAdd{vnn.space@gmail.com}
\emailAdd{eunilwon@korea.ac.kr}

\abstract{We propose a new method to detect gravitational waves, based on spatial coherence interferometry 
with stellar light, as opposed to the conventional temporal coherence interferometry with laser 
sources. The proposed method detects gravitational waves by using two coherent beams 
of light from a single distant star measured at separate space-based 
detectors with a long baseline. This method can be applied to either the amplitude 
or intensity interferometry. This experiment allows for the search of gravitational 
waves in the lower frequency range of $10^{-6}$ to $10^{-4}$ Hz. In this work, we present the 
detection sensitivity of the proposed stellar interferometer by 
taking the detector response and shot and acceleration noises into account. 
Furthermore, the proposed experimental setup is capable of searching for 
primordial black holes and studying the size of the target neutron star, which 
are also discussed in the paper.
}


\arxivnumber{1906.06018}

\begin{document} 
\maketitle
\flushbottom

\section{Introduction}

The discovery of gravitational waves (GWs) that originate from mergers 
of compact objects i.e. black holes (BHs) and neutron stars 
(NSs)~\cite{ref:abbott} has successfully led us to a new era of physics, 
demonstrating Einstein’s theory of General Relativity. The recent detection 
of the GW170817 event by the LIGO and Virgo groups~\cite{ref:abbott2,ref:abbott3}, together 
with the successful electromagnetic (EM) follow-up observations 
by more than 70 observatories, has opened a new opportunity 
to better understand the makings of the universe. In addition, 
multi-wavelength observations will become increasingly important for 
GW astronomy, analogous to EM astronomy over the entire range 
of frequencies that has advanced over the preceding decades.

Multi-wavelength GW observation is foreseen from various 
detection methods. The most successful GW detectors, at present, 
are ground-based laser interferometers, i.e. 
LIGO~\cite{ref:abramovici, ref:harry}, Virgo~\cite{ref:acernese}, 
and KAGRA~\cite{ref:akutsu}, while space-based laser 
interferometers such as LISA~\cite{ref:danzmann}, DECIGO~\cite{ref:kawamura}, 
and  BBO~\cite{ref:crowder}  are 
expected to launch in the late 2020s or later. The ground 
interferometers are sensitive to GWs of 
around 100 Hz that are known to be driven by compact binaries, 
supernovae, and pulsars whereas space-based interferometers can 
detect GWs ranging from 1 Hz to $10^{-4}$ Hz with the origin 
of resolvable supermassive blackhole (SMBH) binaries in the 
scale of $\sim 10^6~\textrm{M}_\odot$~\cite{ref:shannon} (Appendix A).

Interferometry, currently, is the most widely used method 
in detecting GWs that uses coherence of light, a measure 
of correlation between the phases at temporally or spatially 
different points on a wave. Laser based interferometers for GW 
detection such as LIGO and LISA are temporal coherence experiments, 
known as Michelson interferometer. These experiments take 
advantage of a significantly long coherence time of lasers. 
Here, we propose a new method, a stellar interferometry for the 
detection of gravitational waves using spatial coherence interferometry, 
as denoted hereafter by SI (Stellar Interferometry for 
gravitational waves). Instead of using a laser as a source, 
this method will use stellar light as the probe of 
space-time disturbance caused by gravitational waves. This 
method focusses on detecting low-frequency band GWs associated 
with SMBHs. The proposed SI can observe GW frequency range of 
$10^{-4}$ Hz to $10^{-6}$ Hz, corresponding to SMBH binaries of mass between
$10^6~\textrm{M}_\odot$ and 
$10^7~\textrm{M}_\odot$.
This would complement the lower parameter space of LISA in the study 
of GWs using a completely different 
method~\cite{ref:detweiler, ref:verbiest, ref:perera, ref:foster}.

Furthermore, SI can be used as a testbed for other 
interesting physics. These include placing a better 
constraint on the size of the target neutron star 
and searching for the evidence of primordial black holes as a dark 
matter candidate. See Appendix A for more detail.

\section{Model}

\begin{figure}[b]
\centering 
\includegraphics[width=.65\textwidth]{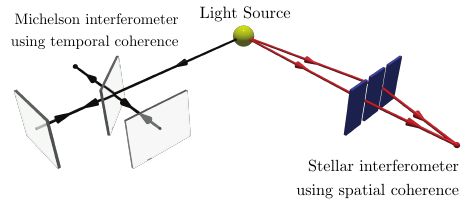}
\caption{\label{fig:interferometer} 
Basic principles of spatial (right) and 
temporal (left) coherence interferometry. Our 
proposed SI uses the spatial coherence while other interferometers 
such as LIGO and LISA use the temporal coherence.
}
\end{figure}

The proposed SI will operate using a stellar source for its 
spatial coherence experiment, while LIGO and other 
similar observatories use lasers as their light source 
for temporal coherence. Basic principles and differences between 
spatial and temporal coherence interferometry are illustrated 
in Fig.~\ref{fig:interferometer}. Stellar interferometry, for 
the purpose of studying stars, was 
first suggested in the 19th century by 
Hippolyte Fizeau~\cite{ref:lawson}. He noted that the diameter 
of an extended disk could be determined interferometrically 
through the measurement of the baseline length at which 
the fringe contrast drops to zero. This concept was first 
exploited by Albert A. Michelson and Francis 
G. Pease~\cite{ref:michelson}. In 1919, Michelson successfully determined the angular 
size of six supergiant stars to milliarcsec scale, among which was
$\alpha$ Orionis (Betelgeuse), measured 
at 0.047 arcsec. Note that for a single star, interference disappears when 
$\ell \geq \ell_s = 1.22~\lambda/\theta_s$ where
$\ell$ is the distance between the two slits, $\lambda$ is the 
wavelength of the incident EM wave, and $\theta_s$ is the angular diameter of the 
star \cite{ref:fizeau}.

\begin{figure}[b]
\centering
\includegraphics[width=0.65\textwidth]{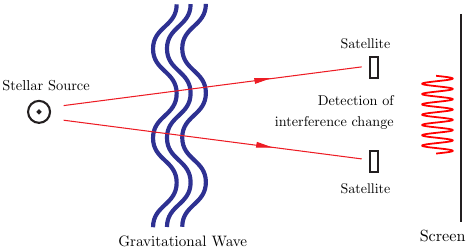}
\caption{\label{fig:methodology} 
Concept of SI detecting GW by measuring changes in interference patterns.
}
\end{figure}

For SI, we apply a similar methodology to detect 
gravitational waves in space. The most fundamental 
concept of SI is depicted in Fig.~\ref{fig:methodology}. While the experimental 
setup for a Michelson stellar interferometer and 
SI are identical, SI will monitor changes in interference patterns 
resulting from disturbances in space caused by gravitational waves.

\begin{figure}[b]
\centering
\includegraphics[width=0.65\textwidth]{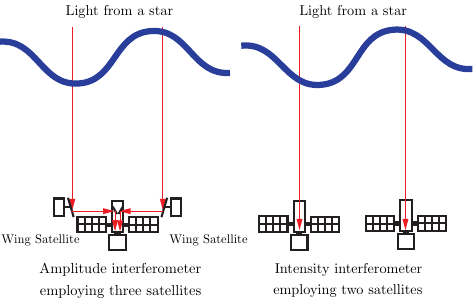}
\caption{\label{fig:satellite} 
A classical method with a three satellites configuration 
(left) and a tandem configuration of two satellites 
(right) for stellar interferometry of detecting gravitational waves.
}
\end{figure}

SI can be best designed to be carried out in space, 
as shown in Fig.~\ref{fig:satellite}. A straightforward configuration 
is a constellation of three satellites (Fig.~\ref{fig:satellite} left). 
A host satellite and two wing satellites, each equipped with 
a telescope for capturing stellar light from a stellar source. 
The two wing satellites will serve as a Michelson stellar 
interferometer by reflecting star light to a host satellite. 
The host satellite can collect and combine the reflected coherent 
lights and the interferometer will look for changes corresponding 
to gravitational waves in the resultant interference patterns.

As an alternative, a two-satellite system configuration 
is also viable, as shown in Fig.~\ref{fig:satellite} right. 
This tandem configuration uses intensity (electrical) 
interferometry, pioneered by Hanbury Brown and 
Twiss~\cite{ref:hanbury1, ref:hanbury2, ref:hanbury3}, 
instead of amplitude (optical) interferometry. 
The intensity interferometer 
is more advantageous in space-based, long baseline 
experiments due to relatively simpler deployment and maneuvering 
of satellites. The two detectors would collect data 
with order picosecond time resolution. All 
of the data obtained from each of the detectors are 
stored that can be analysed offline to measure the 
cross-correlation of the intensity fluctuations 
(second order spatial correlation of electric fields) of 
the stellar light. The data from the two detectors 
with time stamps can be matched to find a pair of two signals 
that have a mutual coherence. This is a fundamental characteristic of intensity 
interferometry with long baseline, a great advantage over
classical techniques involving real-time cross correlation. 
Real-time cross correlation keeps only the final product 
of the correlation and loses the original data from each 
detector, forbidding any further analysis of raw data. 
Moreover, this method removes the need for real-time 
controlling of the satellites with extreme accuracy and there is 
no physical connection between the two detectors.

In both cases, a drag-free system with a test mass
is required. A summary of the acceleration noise and detailed
description of the noise computation can be found 
in Section 3 and Appendix B, respectively.

The ideal orbit of the satellites can be determined 
depending on the chosen stellar source and  the amount of 
reduction in noise. Some of the candidates would be 
the L2 (Lagrange point 2) orbit for the Sun-Earth system 
or a heliocentric orbit, both of which can be adopted with either 
the intensity or the amplitude interferometry. With minimal gravitational 
perturbation, long lifetime of spacecrafts, low temperature, and 
low background environments, the L2 orbit of the Sun-Earth 
system could be preferred for SI. Since the orbital plane around the 
L2 point can be chosen in any orientation, the position of 
the stellar source should not be problematic for the SI experiment. 
Furthermore, the SI experiment requires relatively light-weight satellites 
without a need for high power laser devices. Such aspects would reduce 
related noises as well as the cost and time required for the realization 
of the experiment.

\section{SI Sensitivity}

\subsection{Stellar source and response function}

The distance between the two slits, i.e. satellites, 
is an important parameter for GW detection and the detection 
sensitivity of SI. The separation between two satellites depends on the 
spatial coherence length of a given star $\ell_s = 1.22~\lambda/\theta_s$
and thereby its distance from the satellites and the size of the star. 
For the visible light of 550 nm, a few promising candidates include 
SMC AB8 (12.83 mag, 197,000 Ly distance, $\ell_s = 450$ km) and the 
Crab Pulsar (16.5 mag, 6,523 Ly distance, $\ell_s = 2,070,990$ km).
A very small area of the Sun can also serve as a stellar source 
with $\sim 10$ magnitude and 0.1 milliarcsec of parallax that corresponds to
$15 \times 15~\textrm{m}^2$  area of the Sun's surface, granting
$\ell_s \leq 3$ km. 
For the calculation of the sensitivity,
we chose a half of the actual computed (maximum) 
$\ell_s$ as our $\ell$, 
which would translate to a fringe visibility of approximately 0.6 in coherence 
experiments. Note that the $\ell_s/2$ is not necessarily the most optimal choice.
For given parameters of a stellar interferometer with 
$\ell_s$ and the magnitude of a star, we need to calculate the 
path length difference perturbed by a GW in terms of strain $h$.
This dimensionless quantity gives a fractional change in the path 
length of a photon emitted from a star, as defined by 
$\delta L_c/L_c$ where $L_c$ is the characteristic length that corresponds 
to half the target wavelength of GW ($\lambda_\textrm{GW}$) in the case of SI. The response 
of the detector should be considered in terms of a response function
$\mathcal{R}$
that reflects the geometrical configuration of the 
detector with respect to the GW propagation. This response degrades 
the sensitivity of GW signal ($h_\textrm{signal}$)
hence this factor is divided from the strain noise
($h_\textrm{noise}$) as given by $h_\textrm{signal} \geq h_\textrm{noise}/\sqrt{\mathcal{R}}$.
A detailed calculation of the response function is discussed in Appendix B.
For the calculation of the sensitivity, we initially assumed orthogonal GW propagation 
direction with respect to our line of sight to obtain maximum sensitivity.
Then, for the final estimation of the sensitivity, we took into account the change in the orientation 
due to orbital movements of the detector satellites.

\subsection{Photon statistics and noise}

The noise components arise from the detection process 
in many varieties and forms, that may sit on top of GW signals. 
The sensitivity of SI is determined primarily by two major sources 
of noise. One is shot noise due to the limitation of the 
luminosity of the target star as a light source. The other is 
acceleration noise that is instrument and/or experiment specific. 
Interferometric detectors are limited at high frequencies by the shot noise, 
which arises due to the randomness in the production of photons from their 
source. Shot noise can be calculated using the uncertainty principle:
$\Delta p \cdot \Delta x = \hbar/2$ where $\hbar$ is the reduced Planck 
constant. In terms of the uncertainty in the photon number
($\Delta N_\gamma$) and displacement ($\Delta x$), 
it can be shown that the minimum detectable change by a gravitational wave is given by 
$\Delta x = \delta L_c = h L_c$ and
$\Delta p = \Delta N_\gamma \times 2\pi \hbar/\lambda$.
Each photon from a source arrives at a receiver at random times but with 
an average rate that is proportional to the signal strength. For this type 
of phenomena, the number of events that occur in a given time interval
$\tau$ varies statistically, following a Poisson distribution. For a stellar 
source with optical power $\mathcal{P}$, the average number of photons that arrive within
$\tau$ is given by $\langle N_\gamma \rangle = \mathcal{P}(\lambda /2\pi \hbar c)\tau$ 
and the root-mean-square deviation becomes $\Delta N_\gamma = \sqrt{\langle N_\gamma \rangle}$.
Thus, a minimum detectable change from the shot noise can be written as
\begin{eqnarray}
\delta L_c = \sqrt{\frac{\lambda \hbar c}{8\pi \mathcal{P} \tau}}.
\label{eq:lc}
\end{eqnarray}
Due to the orbital motion of the spacecraft and the relative position 
of a GW source with respect to the orbit, the maximum amplitude 
cannot 
be achieved. It is reduced by a factor of $\sqrt{5}$, when
an average of interferometer response is taken with respect to the entire 
sky~\cite{ref:thorne}. Furthermore, the separation between two satellites
($\ell = \ell_s/2$)
may not be comparable to the wavelength of GW's, hence the response is further 
impeded by an order of
$\lambda_\textrm{GW}/\ell$. 
There is a trade-off between 
$\ell/\ell_s$ and $L_c/\ell$ that a small $\ell$ means 
higher visibility but it also needs to be high enough to be 
comparable to $L_c$ for better sensitivity.

A star of apparent magnitude 8 yields optical power of 
$\mathcal{P}_\textrm{mag=8} \sim 10^{-10}$ W for R-band with $\lambda = 0.64~\mu$m. Note
that the available optical power for LISA is
$\mathcal{P}_\textrm{LISA available} \sim 2\times 10^{-10}$ W.
Finally, the minimum detectable change in SI can be computed using
\begin{eqnarray}
\delta L_\textrm{SI} &=& 
\delta L_c \cdot
\bigg(\frac{\lambda_\textrm{GW}}{\ell}\bigg) 
=
\delta L_c \cdot
\bigg(\frac{4L_c}{\ell_s}\bigg) 
\nonumber \\
&=&
(5.62\times 10^{-13}~\textrm{m}) \cdot
\bigg(
\frac{\lambda}{100~\textrm{nm}}
\bigg)^{\frac{1}{2}}
\nonumber \\
&\cdot&
\big(2.512^{(\textrm{mag}-8)}\big)^{\frac{1}{2}} \cdot
\bigg(
\frac{\tau}{1~\textrm{s}} 
\bigg)^{-\frac{1}{2}}
\cdot \bigg( \frac{ 4 L_c}{\ell_s} \bigg)
\label{eq:deltal}
\end{eqnarray}
where $\tau$ is the integration time equivalent to $L_c/c$. 
This equation includes our choice of 
$\ell_s/2$, which implicitly contains the effect of having the visibility of 0.6.
A full 
derivation of Eq.~(\ref{eq:deltal})  can be found in Appendix B.

Another noise component, dominant in the lower frequency range, 
is the acceleration noise. This noise is associated with external forces and is 
produced from fluctuations of magnetic field, electric field, gravity, temperature, and 
pressure acting on the detector. For SI, the orbital plane of the satellites can 
specifically be chosen to be maintained perpendicular to the line of sight to the 
reference star. SI will have the light propagation paths perpendicular to the satellite 
orbital plane, and thus potential forces, i.e. drag forces. This 
means any associated acceleration noise, in the direction of 
the satellite motion, should not affect SI. In any case, an accurate force 
sensing system with a drag-free system consisting a test mass can measure these acceleration noise and estimate the effects of 
various forces, which will be important for distinguishing them from the effects of 
a true GW signal.

In this work, we separately consider three categories of acceleration noise 
acting on a test mass, 
namely, magnetic field, thermal, and other white noise. 
The test mass for the calculation of acceleration noise 
is made of Pt-Au alloy with a mass of 2.00~kg and a dimension of (4.64~cm)$^3$, similar to 
the one used for the LISA pathfinder~\cite{ref:lisa_pathfinder}.
For the forces associated with magnetic fields, the two leading components acting on a 
test mass are magnetic field fluctuations due to the couplings between: 1. 
The fluctuation in the spacecraft (self-generated) magnetic field and the gradient 
of the spacecraft magnetic field, 2. the fluctuation in the interstellar magnetic 
field (IMF) and the gradient of the spacecraft magnetic field. The noise components are found to be 
$1.7 \times 10^{-18}~\textrm{m s}^{-2}~\textrm{Hz}^{-1/2}$,
and
$5.6 \times 10^{-17}~\textrm{m s}^{-2}~\textrm{Hz}^{-1/2}$ 
at $10^{-4}$ Hz, respectively. See Appendix B for detailed derivations. 
These noise components may be further reduced by using a modern 
state-of-the-art shielding with active cancellation, minimizing electronic parts 
of the detector and current loops, and choosing appropriate orbital geometry.

The thermal noise arises from the temperature difference between the 
walls of the housing that encases the test mass. For SI, the 
two main factors contributing towards the thermal noise are due to: 1. 
different radiation rates between the walls of the housing, 2. the residing gas 
molecules in the housing interacting with the test mass. The outgassing 
of the housing walls, on the other hand, should be negligible compared to the 
other two effects. In this work, we assume a stabilized and well-monitored environment with
$\Delta T \sim 570~\mu$K, giving us the total thermal noise of
$2.0 \times 10^{-16}~\textrm{m s}^{-2}~\textrm{Hz}^{-1/2}$ 
at $10^{-4}$ Hz. Refer to Appendix B for the full derivation.

The remaining noise for SI is white noise of which the 
two most dominant components are due to: in-phase 
transformers and collisions by cosmic-ray particles. The in-phase 
transformer noise is produced from the heating of the transformers 
placed in the drag-free system. On the other hand, the cosmic 
rays produce noise when they collide into the test mass, which 
the results have been calculated by other works. All in all, we 
find the contributions from the two noise components to be
$1.8 \times 10^{-18}~\textrm{m s}^{-2}~\textrm{Hz}^{-1/2}$,
and
$1.2 \times 10^{-18}~\textrm{m s}^{-2}~\textrm{Hz}^{-1/2}$,
respectively (see Appendix B).

\subsection{Results}

\begin{figure}[b]
\centering
\includegraphics[width=0.65\textwidth]{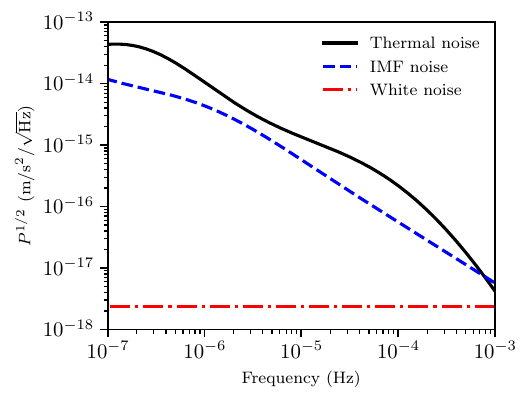}
\caption{\label{fig:noise} 
The power spectral density of each acceleration noise. 
Each line represents 
the thermal noise (black, solid), 
the IMF acceleration noise (blue, dashed), 
and
other white noise (red, dot dashed).
}
\end{figure}

The overall acceleration noise is computed as 
$2.0 \times 10^{-16}~\textrm{m s}^{-2}~\textrm{Hz}^{-1/2}$
at $10^{-4}$ Hz. Each of the acceleration noise component represented 
as power spectral density  
is presented in 
Table~\ref{table:noise} and 
shown in Fig.~\ref{fig:noise}.

\begin{table}
\centering
\begin{tabular}{cc}
\hline
 & Power spectral density \\
Type of acceleration noise & $(\textrm{m s}^{-2}~\textrm{Hz}^{-1/2}$)  \\
&  at $10^{-4}$ Hz\\
\hline
Spacecraft magnetic field effect & $1.7 \times 10^{-18}$ \\
Interplanetary magnetic field effect & $5.6 \times 10^{-17}$ \\
{\bf Combined magnetic field effect} & $5.6 \times 10^{-17}$ \\
Radiation & $7.6 \times 10^{-17}$ \\
Radiometer & $1.2 \times 10^{-16}$ \\
{\bf Combined thermal effect} & $2.0 \times 10^{-16}$ \\
Transformer thermal noise & $1.8 \times 10^{-18}$ \\
Cosmic ray momentum transfer & $1.2 \times 10^{-18}$ \\
{\bf Overall acceleration noise } & $2.0 \times 10^{-16}$ \\
\hline
\end{tabular}
\caption{\label{table:noise}
Estimated acceleration noise components per detector 
in terms of 
power spectral density  
at $10^{-4}$ Hz.
}
\end{table}

\begin{figure*}[b]
\centering
\includegraphics[width=1.00\textwidth]{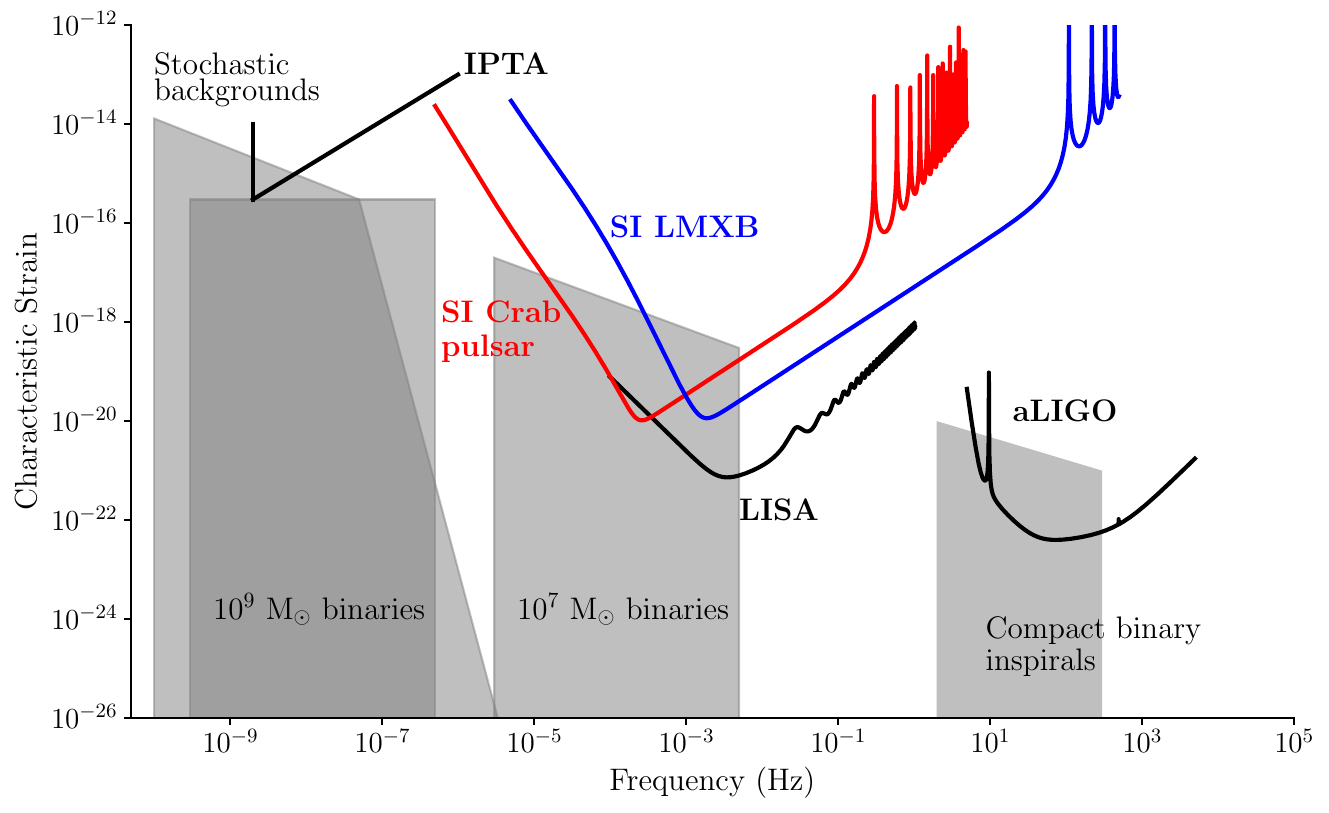}
\caption{\label{fig:sensitivity} 
The sensitivity of our proposed SI experiment compared with other gravitational 
wave experiments currently in operation or planned as future 
experiments. The red curve is using the Crab pulsar as the visible light source and the blue curve is using 4U 1608-522 as the X-ray source. Three solid lines and curves (black) correspond to IPTA sensitivity 
value extracted from Ref.~\cite{ref:moore}, LISA sensitivity data 
from Ref.~\cite{ref:roboson}, and aLIGO sensitivity data from Ref.~\cite{ref:barsotti}, 
from top-left to bottom-right as indicated. The gravitational wave source candidates 
are also indicated with four gray areas~\cite{ref:moore}. Note that the SI curves are 5$\sigma$ curves.
}
\end{figure*}

Figure~\ref{fig:sensitivity} shows the sensitivity budget for SI with 
the shot noise, acceleration noise, and response function included. 
After taking into consideration all the factors, the final strain sensitivity of 
SI can be estimated as
$h_\textrm{SI} = \alpha \sqrt{f S_n}$
where
$S_n = \bigg( \frac{P_\textrm{shot}}{L^2_c} + \frac{P_\textrm{acc}}{(2\pi f)^4 L^2_c}\bigg)$
$\cdot \Big( \frac{1}{\mathcal{R}} \Big)$ 
is the strain noise power spectral density squared, and 
$P_\textrm{shot}^{1/2}$ 
and
$P_\textrm{acc}^{1/2}$ 
are the power spectral densities for shot noise and acceleration noise, respectively.
The power spectral density,
$P_\textrm{shot}^{1/2}$ is given by $\delta L_c$ from
Eq.~(\ref{eq:lc}) and $P_\textrm{acc}^{1/2}$ is estimated as shown above. 
Here, the factor $\alpha = 5\sqrt{10}$ includes the effect of the quadratic sum of two detector noises,
variations of gravitational wave directions with respect to our line of sight due to the orbital motion of the spacecrafts,
and the 5$\sigma$ of the signal to noise ratio.
Choosing 550 nm 
visible light from the Crab Pulsar as the stellar source, and 1,000,000 km as the 
displacement between the two satellites, we calculate the GW detection sensitivity for SI in red 
in terms of characteristic strain and GW frequency as shown in 
Fig.~\ref{fig:sensitivity}.

According to Eq.~(\ref{eq:deltal}), the sensitivity improves when using the 
starlight with shorter wavelengths. From the relation of $\ell_s = 1.22~\lambda/\theta_s$,
however, longer source wavelengths should always be preferred to achieve larger
$\ell_s$ when using the same stellar source with equal angular size. Since 
X-rays have relatively shorter wavelengths, a smaller source and/or higher power is required to
have the sensitivities comparable to those of UV/optical. Low-mass X-ray 
binaries (LMXBs) can be one of the best X-ray source candidates for SI because of 
their relatively high power and the small size of neutron stars. 
For X-rays of 2 keV energy, using a typical neutron star radius of 13 km and a distance 
of 3.6 kpc to 4U 1608-522, the expected satellite separation is of the order 3,200 km. A sensitivity curve for 4U 1608-522 is presented in Fig.~\ref{fig:sensitivity} in blue along with the Crab pulsar in red. Here, we assume 10\% detector efficiency during the computation of the X-ray results.
Furthermore, using X-rays can also lead to other interesting physics studies such as searching for primordial 
black holes as dark matter (DM) candidates and studying macroscopic properties of the 
target neutron star. For more details, please see Appendix A.



\section{Conclusion}

We propose, for the first time, stellar interferometry for the detection 
of gravitational waves with spatial coherence of starlight. We calculate 
the GW detection sensitivity for both stellar sources by considering the shot 
and acceleration noise components and the response function for the stellar 
interferometer. From Fig.~\ref{fig:sensitivity}, we obtain the minimum characteristic strain of
$1.0 \times 10^{-20}$ at $2.5 \times 10^{-4}$ Hz 
for the Crab pulsar, where the acceleration noise is divided into three components, namely 
magnetic, thermal, and white, which in total gives 
$6.2 \times 10^{-17}~\textrm{m s}^{-2}~\textrm{Hz}^{-1/2}$ at $2.5 \times 10^{-4}$ Hz. 
Also, the shot noise is found to be
$3.8 \times 10^{-11}~\textrm{m}~\textrm{Hz}^{-1/2}$.
As well as optical stellar sources, X-rays emitted from LMXBs can be 
used as the probes for gravitational waves. For 4U 1608-522, we obtain 
the minimum characteristic strain of
$1.1 \times 10^{-20}$ at $2.0 \times 10^{-3}$ Hz,
where the acceleration (shot) noise is 
$3.6 \times 10^{-18}~\textrm{m s}^{-2}~\textrm{Hz}^{-1/2}$ 
($4.3 \times 10^{-14}~\textrm{m}~\textrm{Hz}^{-1/2}$).
We anticipate that this 
new method will benefit the field, producing complementary results to further 
our knowledge on gravitational waves.

\appendix

\section{Other applications}

\subsection{Gravitational wave sources of low frequency bands}

Supermassive black holes (SMBHs) are expected to exist at 
the center of almost all galaxies and have a typical mass of around 
$10^5 - 10^{10}~\textrm{M}_\odot$. The observed correlations between the 
mass of SMBHs and the velocity dispersion, mid-infrared luminosity, and mass 
of stellar bulges inside host galaxies imply the importance of SMBHs for 
studying the formation and evolution of galaxies. Recent observations of 
distant quasars indicate that SMBHs have existed from the early 
stages of the universe, 690 Myr after the Big Bang~\cite{ref:banados}. Although the formation 
and evolution of SMBHs are still unclear, it is believed that the SMBHs evolved from 
initial seeds with a mass of around
$10^2 - 10^5~\textrm{M}_\odot$ at a redshift of $10 \leq z \leq 15$.
These initial seeds may have formed from the remnants of Population III 
stars or directly from the collapse of dense gas clouds. In $\Lambda$CDM cosmology, dark 
matter halos and galaxies are formed hierarchically by the merger of smaller 
structures. When two galaxies merge, the SMBHs at the center 
of each galaxy sink into the central region of the merged galaxy 
by dynamical friction and form a binary system. An SMBH binary then 
spirals in and eventually merges as it loses gravitational energy in the 
form of GWs that are detectable at Earth. During an inspiral of 
SMBH binaries, the frequency of GW increases with time, making characteristic tracks on 
the plane of the frequency-characteristic strain. 
With a mass of  
$10^6~\textrm{M}_\odot$, the frequency should span from
$10^{-5}$ Hz to $10^{-3}$ Hz  
with a characteristic strain amplitude of 
$10^{-18}$  for a few months. The proposed stellar interferometer can 
observe SMBH binaries with their total mass spanning between
$10^6~\textrm{M}_\odot$ and 
$10^8~\textrm{M}_\odot$ by focusing on the GW frequency range of
$10^{-6}$ Hz to $10^{-4}$ Hz. 
This fills yet to be considered parameter space in the study of 
gravitational waves, i.e. the gap between the range of LISA
($10^{-2}$ Hz) and the one of PTA ($10^{-9}$ Hz). 
The measurement of the amplitude and spectrum of GWs in the frequency range between
$10^{-6}$ Hz to $10^{-4}$ Hz will enable us to better understand the formation 
and evolution of SMBHs, as well as their link to galaxy evolution. Another target 
GW source of SI frequency band is extreme mass-ratio-inspirals (EMRIs) with a mass ratio greater than 
$10^4$ : 1 between the binary pairs. The large mass difference between two objects 
allows us to observe the effects of gravity in the strong-field limit, 
which is difficult to achieve with stellar-mass binaries. 
EMRIs~\cite{ref:amaro, ref:aharon} are binary systems containing a massive black hole 
that eventually merges with its smaller companion such as a white 
dwarf, neutron star, stellar mass black hole or a giant star with a 
helium core. The final stages of EMRI mergers produce GWs between 
$10^{-4}$ Hz to $10^{-1}$ Hz. LISA is most sensitive in this frequency band, 
while the marginal range of the gradually inspiraling phase can be covered by SI.

\subsection{Probing primordial black hole (PBH) DM with LMXB lensing parallax}

\begin{figure}[b]
\centering
\includegraphics[width=0.65\textwidth]{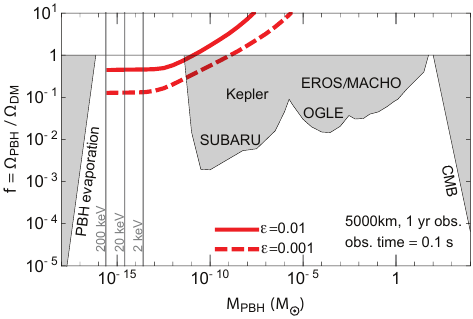}
\caption{\label{fig:pbh} 
The sensitivity of the LMXB lensing parallax on the PBH DM 
abundance, for $1-$year tracking of a LMXB at 6 kpc by 
SI with the baseline 5,000 km. The fractional brightness resolution obtained from 100 ms 
stacking is assumed to be $\epsilon = 0.01$ (solid) and 0.001 (dashed). The typical X-ray spectrum peaks at 
$2-200$ keV, 
which can probe PBHs heavier than the vertical lines. Also shown 
as shaded regions are existing constraints; see~\cite{ref:jung} for more details.
}
\end{figure}

The PBH in the lightest mass window of 
$10^{-16}~\textrm{M}_\odot \sim 10^{-11}~\textrm{M}_\odot$
can theoretically account for the full DM abundance. The only reliable 
lensing method that can probe this window is lensing parallax of a compact source 
such as gamma-ray bursts (GRBs)~\cite{ref:jung, ref:nemiroff}  and microlensing of 
LMXBs~\cite{ref:bai}. SI with long baseline is potentially an ideal laboratory for 
the detection of lensing parallax. In this subsection, we provide a brief 
estimation of LMXB lensing parallax using an SI setup. The Einstein radius 
of the lightest PBH is given by 
$r_E = 7.2 - 2,400$ km for $M_\textrm{PBH} = 10^{-16} - 10^{-11}~\textrm{M}_\odot$ 
at a Gpc distance. This is indeed shorter than the SI baseline 
of 5,000 km so one of the detectors may measure more lensing magnified 
LMXB than the other, i.e. a lensing parallax. The brightness resolution is 
then a key detector parameter. We will assume a fractional 
resolution obtained from 100 ms stacking to be 
$\epsilon = 0.01$ 
(reasonable) or 0.001 (optimistic). With a yearlong tracking of a single LMXB, SI can essentially cover many 
$(10^6 - 10^9)$ patches of the Einstein angle; it is equivalent 
to observing the same large number of LMXBs. The large number can compensate for the 
small lensing optical depth of a nearby LMXB ($\sim$ 6 kpc).
The source of thermonuclear LMXBs is thought to be the neutron star with 
a radius of $\sim$ 10 km, which indeed appears to be smaller than the Einstein angle, 
hence allowing efficient lensing. The X-ray spectrum of LMXBs typically peaks at $2-200$ keV, 
which is high enough to see PBHs more massive than
$10^{-15}~\textrm{M}_\odot$. 
The estimated sensitivity on the PBH DM abundance is shown in 
Fig.~\ref{fig:pbh}. With the aforementioned SI design, a large part 
of the unconstrained PBH mass window can indeed be probed using 
SI. The most crucial capability that can be further improved 
is the brightness resolution, e.g. via larger detector area and/or higher sampling frequency.

\subsection{Probing macroscopic properties of neutron stars}

The two X-ray detectors of SI can be used as an HBT intensity 
interferometer~\cite{ref:hanbury1, ref:hanbury2, ref:hanbury3},
which can measure the radius of a target neutron star. Each detector 
can be used as an X-ray spectrometer that can measure the mass and 
the radius of the neutron star simultaneously. Macroscopic properties of neutron 
stars, such as masses, radii, and temperatures, have been studied using 
X-ray spectroscopy during quiescence of LMXBs~\cite{ref:degenaar}.
Typical temperatures of neutron 
stars in the observed LMXBs are in the range of
$1 - 3$ keV~\cite{ref:ozel, ref:degenaar}. 
With two X-ray detectors of SI that are sensitive to $1 - 10$ keV
photons, one can measure the temperature of the target neutron star. In combination 
with the total measured flux, the mass and radius of the neutron star can be 
estimated by assuming blackbody radiation and the Eddington flux~\cite{ref:degenaar}. This information 
can provide constraints on the neutron star equation of state to determine the inner structure 
of neutron stars. Neutron star equation of state is still unknown mainly due 
to the uncertainties in the high-density behavior of dense hadronic matter. With 
the help of recent measurement of tidal deformability of neutron stars from the 
gravitational wave event GW170817~\cite{ref:abbott2, ref:abbott3}, very hard equations of state have 
been ruled out. On the other hand, observations of 
2 $\textrm{M}_\odot$ 
neutron stars in neutron star - white dwarf binaries ruled out very 
soft equations of state~\cite{ref:demorest, ref:antoniadis, ref:cromartie}. 
Recent observations by the Neutron Star 
Interior Composition Explorer also put new constraints on the radii of 
neutron stars~\cite{ref:riley, ref:miler}. Even though these results narrowed down the region 
of allowed equations of state in the parametric space of neutron star mass and 
radius, the allowed region is still quite large. Therefore, simultaneous measurements 
of mass and radius of a neutron star by SI will produce important results to better 
understand the physics of high dense hadronic matter.

\subsection{Relic stochastic background gravitational waves}

The relic stochastic background of gravitational waves (RSBGW) was predicted 
from zero quantum oscillations due to strong gravitational fields of the 
early universe~\cite{ref:grishchuk1, ref:grishchuk2}. The primary spectrum of RSBGW depends on 
the parameters and models of the expanding universe. The current 
spectrum is a result of the evolution of the Universe, in a 
particular of re-processing the primary spectrum at stages dominated by 
radiation and matter~\cite{ref:lammerzahl, ref:grishchuk}. Indeed, the spectrum can be described 
by a power law. The primary spectrum is intact at frequencies below 
the Hubble frequency of 
$2 \times 10^{-18}$ Hz, and thus, the normalization of 
an observable spectrum can be obtained from the current observations of the CMB.

The sensitivity of the SI at $10^{-7}$ Hz is marginally comparable with the currently assumed amplitude of
$10^{-17}$ 
for RSBGW. The spectrum registering at these frequencies is composed at 
least from a binary stochastic background and RSBGW. It is 
a stochastic signal arising from un-resolvable signals of numerous 
binary systems. The binary background can be estimated much better 
than the RSBGW. Hence, we will have a unique opportunity to directly estimating 
the RSBGW amplitude at the low frequencies. Estimates of upper limits 
of the stochastic gravitational-wave background as well as the 
binary stochastic background have already been obtained by LIGO/Virgo 
observations at high frequencies~\cite{ref:abbott4, ref:abbott5}.

\section{Computing SI Sensitivity}

\subsection{Full Derivation of characteristic strain}

\begin{figure}[b]
\centering
\includegraphics[width=0.65\textwidth]{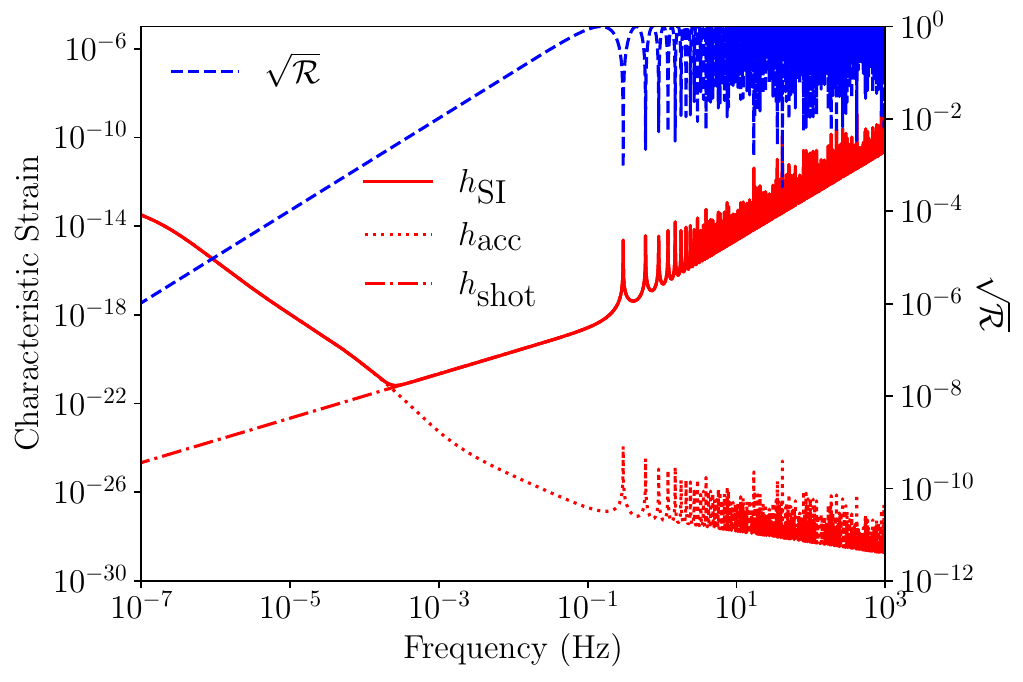}
\caption{\label{fig:response} 
Sensitivity of SI in terms of characteristic strain ($h_c$) for the 
stellar source of 550 nm visible light from Crab Pulsar (red).
The dot dashed (dotted) curve 
represents the sensitivity of SI corresponding to shot (acceleration) noise. 
Square-root of the response functions are also shown (dashed).
}
\end{figure}

This derivation follows closely the procedures shown in~\cite{ref:maggiore}, and was altered to 
conform with the experimental setup of SI. The average power of 
starlight during the observation time
$\tau$ is given by $\mathcal{P} = \langle N_\gamma \rangle\hbar \omega/\tau$ where
$\langle N_\gamma \rangle$ is the average number of photons arriving at the detector within $\tau$, 
$\omega$ is the angular frequency of the EM wave we will be observing. Using Poisson distribution and 
$\Delta N_\gamma = \sqrt{ \langle N_\gamma \rangle}$, we get:
\begin{eqnarray}
\Delta \mathcal{P} 
= \frac{1}{\tau} \sqrt{\langle N_\gamma \rangle} \hbar \omega 
= \sqrt{\frac{\hbar \omega \mathcal{P}}{\tau}}.
\label{eq:deltap1}
\end{eqnarray}
For the GW propagating in the $z$-direction, 
the space-time interval is given by
\begin{eqnarray}
ds^2 &=& -c^2 dt^2 + [ 1+ h_+(t,z)] dx^2  + [1-h_+ (t,z)] dy^2 
\nonumber \\
&+& dz^2,
\end{eqnarray}
where we used ``+ polarization'' 
only for simplicity given by $h_+(t,z) = h_0 \cos (\omega_\textrm{GW} t - k_\textrm{GW} z)$.
For the starlight travelling towards the satellites in the $x$-axis, we get
\begin{eqnarray}
dx = c \Big[ 1- \frac{1}{2} h_+(t)\Big]dt.
\label{eq:dx}
\end{eqnarray}
Also, the two signals for the two satellites separated 
by $\ell$ in the $z$-axis (the satellite (1) is located at $z=-\ell$ and (2) is at $z=0$), are
given respectively by,
\begin{eqnarray}
h_{(1)} &\equiv& h_+(t, -\ell) = h_0 \cos{(\omega_\textrm{GW}t + k_\textrm{GW}}\ell),
\nonumber \\
h_{(2)} &\equiv& h_+(t,0) = h_0 \cos{(\omega_\textrm{GW}t)},
\label{eq:two_satellites}
\end{eqnarray}
where $h_0$ and $\omega_\textrm{GW}$ ($k_\textrm{GW}$) 
are the amplitude and frequency 
(wave number) of the gravitational wave. 
Let us assume photons emitted from our star 
by a gravitational wave intersecting the 
photon propagation path. 
The photon is affected by the gravitational wave from a time $t_0$ and travels a distance $L_c$ at a time $t_1$.
Using Eq.~(\ref{eq:dx}), the distance $L_c$ becomes:
\begin{eqnarray}
L_c = c(t_1 - t_0) - \frac{c}{2} \int^{t_1}_{t_0} h_+ (t^\prime,z)~dt^\prime.
\end{eqnarray}
To first order, $t_1 = t_0 + L_c/c$ hence at $z=- \ell$ we can re-writte $t_1-t_0$ as:
\begin{eqnarray}
t_1 -t_0 &=& \frac{L_c}{c} 
+\frac{1}{2} \int^{t_0 + \frac{L_c}{c}}_{t_0} 
h_0 \cos{(\omega_\textrm{GW} t^\prime + k_\textrm{GW} \ell)}~dt^\prime 
\nonumber \\
&=& \frac{L_c}{c} 
+\frac{h_0 L_c}{2c} \frac{\sin{(\omega_\textrm{GW} L_c/2c)}}{\omega_\textrm{GW} L_c/2c}
\nonumber \\
&&\qquad \times \cos{\big[\omega_\textrm{GW} (t_0 +L_c/2c) + k_\textrm{GW} \ell\big]}.
\end{eqnarray}
Since we observe the photon at a given time $t$,  now setting 
$t= t_1$ and using Eq.~(\ref{eq:two_satellites}), we get the time $t_0$ 
\begin{eqnarray}
t_0 &=& t - \frac{L_c}{c}
\nonumber \\
&-& \frac{L_c}{2c} h\bigg( t_0 + \frac{L_c}{2c}, -\ell \bigg)
\frac{\sin{(\omega_\textrm{GW} L_c/2c)}}{\omega_\textrm{GW} L_c/2c}. 
\label{eq:t0}
\end{eqnarray}
Using sinc function and $t_0 +L_c/2c \simeq t-L_c/2c $ at leading order,
we can rewrite Eq.~(\ref{eq:t0})  for the first satellite as:
\begin{eqnarray}
t_{0,(1)} &=& t - \frac{L_c}{c}  
\nonumber \\
&-& \frac{L_c}{2c} h\bigg( t-\frac{L_c}{2c}, -\ell \bigg)
\textrm{sinc}{(\omega_\textrm{GW} L_c/2c)}.
\label{eq:sinc}
\end{eqnarray}
The last term in Eq.~(\ref{eq:sinc}) is the phase change due to gravitational wave 
for the first satellite, denoted as
$\Delta \phi_1/\omega$. Similarly, for the second satellite located at $z=0$, we follow the same 
procedure with the second equation of Eq.~(\ref{eq:two_satellites}). 
Hence, we simply remove $\ell$ from Eq.~(\ref{eq:sinc}) and get:
\begin{eqnarray}
t_{0,(2)} &=& t - \frac{L_c}{c}  
\nonumber \\
&-& \frac{L_c}{2c} h\bigg( t-\frac{L_c}{2c},0 \bigg)
\textrm{sinc}{(\omega_\textrm{GW} L_c/2c)}.
\end{eqnarray}
Therefore, the phase difference of two lights measured at the satellites (1) and  (2)
\begin{eqnarray}
\Delta \phi_\textrm{SI}(t) &=& \Delta \phi_1 - \Delta \phi_2
\nonumber \\
&=&
\frac{L_ch_0}{c} \omega \textrm{sinc}(\omega_\textrm{GW} L_c/2c)
\sin{(k_\textrm{GW} \ell/2)}
\nonumber \\
&\times&
\sin{\big[ \omega_\textrm{GW}(t-L_c/2c)+ k_\textrm{GW} \ell/2\big]}.
\end{eqnarray}
Here we define the response function $\mathcal{R}$ as
\begin{eqnarray}
\mathcal{R} \equiv \textrm{sinc}(\omega_\textrm{GW} L_c/2c)
\sin{(k_\textrm{GW} \ell/2)}
\end{eqnarray}
up to  the time-dependent term. For the 
low frequencies of GW, the response function is reduced by 
the second term representing the ratio between the spacing between 
the two satellites and the wavelength of the gravitational wave 
being detected. On the other hand, for the high frequencies, the 
response function oscillates. 
Fig.~\ref{fig:response} shows the response function for the Crab Pulsar 
observed with 550 nm visible light.
Here, we specifically picked the observation 
time so that $L_c= c\tau = \lambda_\textrm{GW}/2$. 
In the case of the distance from the star to the satellites being 
much larger than $L_c$, the effects of the gravitational waves 
corresponding to the integer multiples of 
$\lambda_\textrm{GW}$ will be cancelled out.
Now, for the total signal 
power~\cite{ref:maggiore}, we have
\begin{eqnarray}
\mathcal{P}_\textrm{GW} = \mathcal{P}\sin^2{(\phi)},
\end{eqnarray}
and the variation in power due to a GW is 
\begin{eqnarray}
\Delta \mathcal{P}_\textrm{GW} &=& \frac{\mathcal{P}}{2} |\sin{(2\phi_0)}|  \Delta \phi_\textrm{SI}
\end{eqnarray}
where we used 
$\Delta \phi_1 = - \Delta \phi_2$
for the simplest case,
and
the phase  $\phi_0$ is a parameter that the experimenter 
can adjust, hence we can choose the best working point.
The variation in power due to the shot noise is
\begin{eqnarray}
\Delta \mathcal{P}_{\rm shot} =\Delta \mathcal{P} |\sin\phi_0|=  \sqrt{\frac{\hbar \omega \mathcal{P} }{\tau}} |\sin\phi_0|.
\end{eqnarray}
Finally, we obtain the signal-to-noise ratio:
\begin{eqnarray}
\frac{S}{N} &=& \frac{\Delta \mathcal{P}_\textrm{GW}}{\Delta \mathcal{P}_{\rm shot}}
=  \sqrt{\frac{\tau \mathcal{P}}{\hbar \omega}} |\cos\phi_0| \Delta \phi_{\rm SI}.
\end{eqnarray}
For the characteristic length $L_c = c\tau = \lambda_\textrm{GW}/2$, 
and in the limit of $k_{\rm GW} \ell \ll 1$, and $\cos
\phi_0 = \frac{1}{\sqrt2}$, we obtain
\begin{eqnarray}
\frac{S}{N} &=& \sqrt{\frac{\tau \omega\mathcal{P}  }{2\hbar}} \frac{h_0 \ell }{c}.
\end{eqnarray}
Combining above with the signal-to-noise ratio 
equation written in terms of the strain sensitivity
$S^{1/2}_n (f)$~\cite{ref:maggiore, ref:moore2},
\begin{eqnarray}
\frac{S}{N} = \bigg[ \frac{\tau}{S_n(f)}\bigg]^{1/2} h_0
\end{eqnarray}
we find that
\begin{eqnarray}
\sqrt{S_n(f)} = \sqrt{\frac{2\hbar}{\omega \mathcal{P}}} \frac{c}{\ell}.
\end{eqnarray}
Hence, in Fig.~\ref{fig:sensitivity}, we plot the characteristic strain defined as~\cite{ref:moore}:
\begin{eqnarray}
h_c (f) &=& 
\sqrt{f S_n(f)} = \sqrt{\frac{\lambda \hbar c}{2\pi \mathcal{P} \tau}} \frac{1}{\ell} 
\end{eqnarray}
where $f = 1/(2\tau)$, 
which indeed is the same as what we obtained in Eq.~(\ref{eq:deltal}) 
from the uncertainty principle.
In Fig.~\ref{fig:response},  
$h_c(f)$
are shown again with the sensitivity of SI only with the shot noise or the acceleration noise.

\subsection{Magnetic field effect}

 The magnetic fields affecting the SI are due to ``spacecraft'' ($B_\textrm{SC}$) 
and interplanetary ($B_\textrm{IMF}$).
The amount of force, due to these magnetic fields, experienced by the test mass is given by  
$F = \frac{\chi V}{\mu_0} [
B_\textrm{SC} \cdot \nabla B_\textrm{SC} +$
$ B_\textrm{SC} \cdot \nabla B_\textrm{IMF} +$
$ B_\textrm{IMF} \cdot \nabla B_\textrm{SC} +$
$ B_\textrm{IMF} \cdot \nabla B_\textrm{IMF}]$,
where $V$ is the volume of the test mass, $\chi$ is its magnetic susceptibility, and
$\mu_0$ is the vacuum permeability. Since the gradient of
$B_\textrm{IMF}$ is very small (0.5 nT/10,000 km), $\nabla B_\textrm{IMF}$ 
terms are ignored. Here, we assume a Pt-Au alloy test mass of 2 kg with
$\chi = 1.7 \times 10^{-5}$ and test mass density 
$\rho_\textrm{TM} = 2.0 \times 10^{4}~\textrm{kg m}^{-3}$. For $B_\textrm{IMF}$, data from
reference~\cite{ref:filloux} have been used. Hence, the power spectral 
density squared on the magnetic field effects at 
$10^{-4}$ Hz is
\begin{eqnarray}
P_\textrm{magnetic} &=&
\Bigg[
\frac{\chi}{\rho_\textrm{TM} \mu_0} B_\textrm{SC} \cdot
\bigg(
\frac{\partial B_\textrm{SC}}{\partial x_\textrm{SC}}
\Big)
\Bigg]^2 
\nonumber \\
&+&
\Bigg[
\frac{\chi}{\rho_\textrm{TM} \mu_0} B_\textrm{IMF} \cdot
\bigg(
\frac{\partial B_\textrm{SC}}{\partial x_\textrm{SC}}
\bigg)
\Bigg]^2 
\end{eqnarray}
where we assume $B_\textrm{SC}$ and $\partial B_\textrm{SC}/\partial x_\textrm{SC}$ to be
$5.0 \times 10^{-8}~\textrm{T Hz}^{-1/2}$ and
$5.0 \times 10^{-8}~\textrm{T m}^{-1}$, respectively. From this equation, 
$P^{1/2}_\textrm{magnetic} = 5.6 \times 10^{-17}~\textrm{m s}^{-2}~\textrm{Hz}^{-1/2}$. 

\subsection{Thermal effect}

We have considered three effects caused by pressure changes within the test 
mass chamber due to temperature fluctuations, which are due to outgassing 
and radiation from the chamber walls, and any residing gas within 
the chamber. Firstly, the outgassing rate decreases exponentially with temperature. Hence, 
at 50 K, which is the environment we are considering for this experiment, the 
outgassing rate becomes negligible. Secondly, the noise from the radiation effect arises when the pressure
$p_r = \frac{4\sigma}{3c} T^4$ is produced by the difference in temperature $\Delta T$
between opposite housing walls, i.e. $|T_\textrm{w2} - T_\textrm{w1}| = \Delta T$, where
$\sigma$ is the Stefan-Boltzmann constant. 
Therefore, the change in pressure is given by
\begin{eqnarray}
\Delta p_r &=& \frac{4\sigma}{3c} |T^4_\textrm{w2} - T^4_\textrm{w1}|
\nonumber \\
&\sim&
\frac{4\sigma}{3c} 
\Bigg\{
T^4_\textrm{w1} \bigg(
1+ \frac{4 \Delta T}{T_\textrm{w1}}
\bigg)
- T^4_\textrm{w1}
\Bigg\}
\nonumber \\
&=& \frac{16 \sigma T^3_\textrm{w1}}{3c} \Delta T.
\end{eqnarray}
At $10^{-4}$ Hz range, we expect to be able to control 
$\Delta T$ to 570 $\mu$K. Hence, the acceleration caused by the radiation effect becomes 
\begin{eqnarray}
a_\textrm{radiation} &=& \frac{16 \sigma A T^3_\textrm{w1}}{3c M_\textrm{TM}} \Delta T
\nonumber \\
&\sim& 7.6 \times 10^{-17}~\textrm{m s}^{-2},
\end{eqnarray}
where $A$ is the area of a test mass wall and $M_\textrm{TM}$ is the mass of the test mass.
Lastly, the thermal effect from residual (ideal) gas that 
collide with the test mass should produce acceleration equal to
\begin{eqnarray}
a_\textrm{radiometer} &=& \frac{nA k_B \Delta T}{M_\textrm{TM}} =
\frac{p_{r,0} A \Delta T}{M_\textrm{TM} T_\textrm{w1}}
\nonumber \\
&\sim& 1.2 \times 10^{-16}~\textrm{m s}^{-2},
\end{eqnarray}
where $n$ is the number of moles and $k_B$ is the Boltzmann constant. Here,
$p_{r,0}$ is the pressure within the housing, that when exposed to the L2 space environment 
could be assumed as 
$10^{-8}$ Pa. All in all, the corresponding thermal acceleration noise at 
$10^{-4}$ Hz is
\begin{eqnarray}
P^{1/2}_\textrm{thermal} &\sim&
\Bigg(
\frac{16\sigma AT^3_0}{3 c M_\textrm{TM}} + 
\frac{ p_{r,0} A}{M_\textrm{TM} T_0} 
\Bigg)
P^{1/2}_{\Delta T} (w)
\nonumber \\
&\sim& 2.0 \times 10^{-16}~\textrm{m s}^{-2}~\textrm{Hz}^{-1/2},
\end{eqnarray}
where we assume $T_\textrm{w1} = T_0 = 50$ K
and $P^{1/2}_{\Delta T} (w)$ is the frequency dependent
term after the Fourier transformation of the accelerations above.
The thermal component shown in Fig.~\ref{fig:noise} is
obtained by convolving the solar flux~\cite{ref:solar_flux}
with a transfer function of solar energy from solar panel to detector 
on board the spacecraft.

\subsection{Other noise}
There are two other dominant noise components we consider for this work. First is 
due to the temperature rise of the in-phase transformers in the drag free 
system. Assuming we will use a similar drag free system as LISA, we adopt their value of
$1.8 \times 10^{-18}~\textrm{m s}^{-2}~\textrm{Hz}^{-1/2}$~\cite{ref:bender}.
Second is due to cosmic-ray particles colliding with the test mass, transferring 
their momenta. This can be estimated as 
\begin{eqnarray}
P_\textrm{CR} \sim \frac{2p^2 \mu}{M^2_\textrm{TM}}.
\end{eqnarray}
Here, $p$ is the average momentum of each cosmic-ray particle, 
$\mu$ is the number of particles per time that fully transfer their 
momenta. Considering the incident angle (0.9 sr) and the test mass surface area
($2.1 \times 10^{-3}~\textrm{m}^2$) long with the known cosmic-ray 
flux at 1 GeV, the corresponding cosmic-ray noise is
$P^{1/2}_\textrm{CR} \sim 1.2 \times 10^{-18}~\textrm{m s}^{-2}~\textrm{Hz}^{-1/2}$.

\acknowledgments

We are grateful to Antoine Labeyrie for his suggestion of Crab Pulsar as 
a light source for stellar interferometery. We would also like to thank 
Sunkee Kim, Hyungwon Lee, Jongmann Yang, Soomin Jeong, Chanyeol Kim, 
Chunglee Kim, Jiwoo Nam, Jean Schneider, Denis Mourard, and Rijuparna Chakraborty 
for fruitful discussions on noises and light sources. We acknowledge the support 
from the National Research Foundation (NRF) of Korea: 
I.H.P. NRF-2017K1A4A3015188, NRF-2021R1A2B5B03002645, and NRF-2019H1D3A2A02060090;
K.-Y.C. NRF-2019R1A2B5B010701
81; 
J.H. NRF-2021R1A2C101109811;
S.J.  NRF-2019R1 C1C1010050; 
D.H.K.  NRF-2018R1D1
A1B07051276; 
C.H.L. NRF-2016R1A5A1013277 and NRF-2018R1D
1A1B07048599; 
S.H.O. NRF-2019R1A2C2006787; 
S.C.P.  NRF-2021R1A4A20 01897 and NRF-2019R1A2C1089334; 
C.D.R.  NRF-2018R1A6A1A06024977; 
E.W. NRF-2017 R1A2B3001968. 
A.P. acknowledges the support from RSCF grant 18-12-00378.



\begin{thebibliography}{99}

\bibitem{ref:abbott} 
B. P. Abbott {\it et al.}, {\it Observation of Gravitational Waves from a Binary Black Hole Merger},
Phys. Rev. Lett {\bf 116}, 061102 (2016).

\bibitem{ref:abbott2} 
B. P. Abbott {\it et al.}, {\it Multi-messenger Observations of a Binary Neutron Star Merger},
Astrophys. J. Lett. {\bf 848}, L12 (2017).

\bibitem{ref:abbott3} 
B. P. Abbott {\it et al.}, {\it Observation of Gravitational Waves from a Binary Neutron Star Inspiral},
Phys. Rev. Lett. {\bf 119}, 161101 (2017).

\bibitem{ref:abramovici}
A. Abramovici {\it et al.}, 
{\it LIGO: The Laser Interferometer Gravitational-Wave Observatory}, Science {\bf 256}, 325-333 (1992).

\bibitem{ref:harry}
G. M. Harry {\it et al.}, {\it Advanced LIGO: the next generation of gravitational 
wave detectors}, Classical Quantum Gravity {\bf 27}, 08406 (2010).

\bibitem{ref:acernese}
F. Acernese {\it et al.}, {\it Advanced Virgo: a second-generation interferometric gravitational 
wave detector}, Classical Quantum Gravity {\bf 32}, 024001 (2014).

\bibitem{ref:akutsu}
T. Akutsu {\it et al.}, {\it KAGRA: 2.5 generation interferometric gravitational 
wave detector}, Nature Astronomy {\bf 3}, 35 (2019).

\bibitem{ref:danzmann}
K. Danzmann, {\it LISA Laser Interferometer Space Antenna–A 
proposal in response to the ESA call for L3 mission concepts}, 
Albert Einstein Inst. Hanover, Leibniz Univ. Hanover, Max Planck 
Inst. Gravitational Phys., Hannover, Germany, Tech. Rep (2017).

\bibitem{ref:kawamura}
S. Kawamura {\it et al.}, {\it The Japanese space gravitational wave 
antenna: DECIGO}, Classical Quantum Gravity {\bf 28}, 094011 (2011).

\bibitem{ref:crowder}
Crowder, and N. J. Cornish, 
{\it Beyond LISA: Exploring Future Gravitational Wave Missions}, 
Phys. Rev. D {\bf 72}, 083005 (2005).

\bibitem{ref:shannon}
R. M. Shannon {\it et al.}, 
{\it Gravitational-Wave Limits from Pulsar 
Timing Constrain Supermassive Black Hole Evolution}, Science {\bf 342}, 334 (2013).

\bibitem{ref:detweiler}
S. Detweiler, 
{\it Pulsar timing measurements and the 
search for gravitational waves}, Astrophys. J. {\bf 234}, 1100 (1979).

\bibitem{ref:verbiest}
J. P. W. Verbiest {\it et al.}, 
{\it The International Pulsar Timing Array: 
first data release}, Mon. Not. R. Astron. Soc. {\bf 458}, 1267 (2016).

\bibitem{ref:perera}
B. B. P.  Perera {\it et al.}, 
{\it The International Pulsar Timing Array: second 
data release}, Mon. Not. R. Astron. Soc. {\bf 490}, 4666(2019).

\bibitem{ref:foster}
R. S. Foster, and D. C. Backer, 
{\it Constructing a pulsar timing array}, Astrophys. J. {\bf 261}, 300 (1990).

\bibitem{ref:lawson}
P. R. Lawson, 
{\it Principles of long baseline stellar interferometry}, 
Jet Propulsion Laboratory (California Institute of Technology, 2000).

\bibitem{ref:michelson}
A. A. Michelson, and F. G. Pease, 
{\it Measurement of the Diameter of a Orionis with 
the Interferometer}, Astrophys. J. {\bf 53}, 249 (1921).

\bibitem{ref:fizeau}
H. Fizeau, {\it Prix Bordin: rapport sur le concours de l’annee 1867},
Compt. Rend. Acad. Sci. Paris, 932-934 (1867);
M. Born and E. Wolf, {\it Principle of Optics}, New York: Macmillan, (1964).

\bibitem{ref:hanbury1}
B. R. Hanbury, and R. Q. Twiss, 
{\it Correlation between photons in two coherent beams of light}, Nature {\bf 177}, 27 (1956).

\bibitem{ref:hanbury2}
B. R. Hanbury, and R. Q. Twiss, 
{\it A Test of a New Type of Stellar Interferometer on Sirius}, Nature {\bf 178}, 1046 (1956).

\bibitem{ref:hanbury3}
B. R. Hanbury, and R. Q. Twiss, 
{\it Interferometry of the Intensity Fluctuations in light. I. 
Basic Theory: the Correlation between Photons in Coherent Beams of Radiation},
Proc. R. Soc. A: {\bf 242}, 300 (1957).

\bibitem{ref:thorne}
K. S. Thorne, 
{\it Three Hundred Years of Gravitation}, 
edited by S. W. Hawking and W. Israel (Cambridge University Press, 1987).

\bibitem{ref:lisa_pathfinder}
Armano, M. {\it et al.},
{\it Sub-Femto-$g$ Free Fall for Space-Based Gravitational Wave Observatories: LISA Pathfinder Results},
Phys. Rev. Lett. {\bf 116}, 231101 (2016).

\bibitem{ref:moore}
C. J. Moore, R. H. Cole, and C. P. L. Berry, 
{\it Gravitational-wave sensitivity curves},
Classical Quantum Gravity {\bf 32}, 015014 (2015).

\bibitem{ref:roboson}
T. Roboson, N. Cornish, and C. Liug, 
{\it The construction and use of LISA sensitivity curves}, 
Classical Quantum Gravity {\bf 36}, 105011 (2019).

\bibitem{ref:barsotti}
L. Barsotti, S. Gras, M. Evans, and P. Fritschel, 
{\it The updated Advanced LIGO design curve}, 
LIGO-T1800044-v5 (2018), \url{https://dcc.ligo.org/LIGO-T1800044/public/}.

\bibitem{ref:banados}
E. Ba\~{n}ados et al., {\it An 800-million-solar-mass black hole in a significantly 
neutral Universe at a redshift of 7.5}, Nature 553, 473 (2018).

\bibitem{ref:amaro}
P. Amaro-Seoane, and M. Preto, 
{\it The impact of realistic models of mass segregation on the event 
rate of extreme-mass ratio inspirals and cusp re-growth}, 
Classical Quantum Gravity {\bf 28}, 094017 (2011).

\bibitem{ref:aharon}
D. Aharon, and B. P. Hagai, 
{\it The Impact Of Mass Segregation And Star Formation On The Rates 
Of Gravitational-Wave Sources From Extreme Mass Ratio Inspirals}, 
Astrophys. J. Lett. {\bf 830}, L1 (2016).

\bibitem{ref:jung}
S. Jung, and T. H. Kim, 
{\it GRB lensing parallax: Closing the primordial black hole 
dark matter mass gap}, Phy. Rev. Research 2, 013113 (2020). 

\bibitem{ref:nemiroff}
R. J. Nemiroff, and G. Andrew, 
{\it Probing MACHOs of mass  with gamma-ray burst parallax spacecraft}, 
Astrophys.  J. Lett. {\bf 452}, L111 (1995).

\bibitem{ref:bai}
Y. Bai, and O. Nicholas, 
{\it Microlensing of X-ray Pulsars: a Method to Detect Primordial Black 
Hole Dark Matter}, Phys. Rev. D {\bf 99}, 123019 (2019).

\bibitem{ref:degenaar}
N. Degenaar {\it et al.}, 
{\it Further X-ray observations of EXO 0748-676 in quiescence: evidence 
for a cooling neutron star crust}, Mon. Not. R. Astron. Soc. {\bf 412}, 1409 (2011).

\bibitem{ref:ozel}
F. \"{O}zel, D. Psaltis, T. Güver, G. Baym, C. Heinke, and S. Guillot {\it et al.}, 
{\it The dense matter equation of state from neutron star radius and mass measurements}, 
Astrophys. J. {\bf 820}, 25 (2016).

\bibitem{ref:demorest}
P. B. Demorest, T. Pennucci, S. M. Ransom, M. S. E. Roberts, and J. W. T. Hessels, 
{\it A two-solar-mass neutron star measured using Shapiro delay}, 
Nature {\bf 467}, 1081 (2010).

\bibitem{ref:antoniadis}
J. Antoniadis {\it et al.}, 
{\it A massive pulsar in a compact relativistic binary}, Science {\bf 340}, 1233232 (2013). 

\bibitem{ref:cromartie}
H. T. Cromartie {\it et al.}, 
{\it Relativistic Shapiro delay measurements of an extremely massive millisecond 
pulsar}, Nature Astronomy {\bf 4}, 72 (2019).

\bibitem{ref:riley}
T. D. Riley {\it et al.}, 
{\it A NICER View of PSR J0030+0451: Millisecond Pulsar Parameter Estimation},
Astrophys. J. Lett. {\bf 887}, L21 (2019). 

\bibitem{ref:miler}
M. C. Miller {\it et al.}, 
{\it PSR J0030+0451 Mass and Radius from NICER Data and Implications for 
the Properties of Neutron Star Matter}, 
Astrophys. J. Lett. {\bf 887}, L24 (2019). 

\bibitem{ref:grishchuk1} 
L. P. Grishchuk, 
{\it Amplification of Gravitational Waves in the Isotropic World}, 
Zh. \`{E}ksp. Teor. Fiz {\bf 67}, 825 (1974).

\bibitem{ref:grishchuk2} 
L. P. Grishchuk, {\it Primordial gravitons and possibility of their observation}, 
Zh. \`{E}ksp. Teor. Fiz {\bf 23}, 326 (1976).

\bibitem{ref:lammerzahl} 
C. L\"{a}mmerzahl, C. W. F. Everitt, and F. W. Hehl, 
{\it Gyros, Clocks, Interferometers: Testing Relativistic Gravity in 
Space}, Lecture Notes in physics. vol. {\bf 562}, p167. (Berlin: Springer, 2001).

\bibitem{ref:grishchuk} 
L. P. Grishchuk, in {\it Astrophysics Update}, edited by J. Mason. p281. (Berlin: Springer, 2004).

\bibitem{ref:abbott4} 
B. P. Abbott {\it et al.}, 
{\it Search for the isotropic stochastic background using data from Advanced LIGO’s second observing run}, 
Phys. Rev. D {\bf 100}, 061101 (2019).

\bibitem{ref:abbott5} 
B. P. Abbott {\it et al.}, 
{\it GW170817: implications for the stochastic gravitational-wave background 
from compact binary coalescences}, Phys. Rev. Lett. {\bf 120}, 091101 (2018).

\bibitem{ref:maggiore} 
M. Maggiore, {\it Gravitational Waves Volume 1 : Theory and Experiments}, 
(Oxford university press, 2008).

\bibitem{ref:moore2} 
C. J. Moore, R. H. Cole, and C. P. L. Berry, 
{\it Gravitational-wave sensitivity curves}, Classical Quantum Gravity  {\bf 32}, 015014 (2015).

\bibitem{ref:filloux} 
J.H. Filloux, {\it Instrumentation and experimental methods for oceanic studies}, 
In “Geomagnetism”, ed. J.A. Jacobs, Academic Press, London, pp. 143-248 (1987).

\bibitem{ref:solar_flux}
Fr\"ohlich, Claus, and J. Lean, {\it Solar radiative output and its variability: evidence and mechanisms},
The Astronomy and Astrophysics Review 12.4, 273-320 (2004).

\bibitem{ref:bender} 
P.  Bender {\it et al.}, {\it LISA Pre-Phase A Report}, 
Max-Planck Institut f\"{u}r Quantenoptik (1998).





\end{thebibliography}
\end{document}